\documentclass{mem}
\usepackage{natbib}\usepackage{txfonts}\usepackage{balance}
\usepackage{graphicx}
\usepackage[a4paper]{hyperref}
\idline{75}{1}
\begin{document}
\def\teff{$T\rm_{eff }$}
\def\kms{$\mathrm {km s}^{-1}$}

\title{Rings, spirals and manifolds}
\subtitle{}
\author{E. \,Athanassoula\inst{1}, 
M. \,Romero-G\'omez\inst{1}
\and J. J. \, Masdemont\inst{2}  }

  \offprints{E. Athanassoula}

\institute{
Laboratoire d'Astrophysique de Marseille (LAM), UMR6110, Observatoire
Astronomique de Marseille Provence (OAMP), CNRS/Universit\'e de Provence,
Technop\^ole de Marseille Etoile, 38 rue Fr\'ed\'eric Joliot Curie, 13388 Marseille C\'edex 20, France.\\
\email{lia@oamp.fr, merce.romerogomez@oamp.fr} 
\and
I.E.E.C \& Dep. Mat. Aplicada I, Universitat Polit\`ecnica de
Catalunya, Diagonal 647, 08028 Barcelona, Spain.
}

\authorrunning{Athanassoula et al. }

\titlerunning{Rings, spirals and manifolds}

\abstract{
  Two-armed, grand design spirals and inner and outer rings in barred
  galaxies can be due to orbits guided by the manifolds emanating from
  the vicinity of the $L_1$ and $L_2$ Lagrangian points, located at
  the ends of the bar. We first summarise the necessary theoretical
  background and in particular we describe the dynamics around the
  unstable equilibrium points in 
  barred galaxy models, and the corresponding homoclinic and
  heteroclinic orbits. We then discuss two specific morphologies and the
  circulation of material within the corresponding manifolds. We also
  discuss the case where 
  mass concentrations at the end of the bar can stabilise the $L_1$
  and $L_2$ and the relevance of this work to the gas concentrations
  in spirals and rings.  

\keywords{galaxies: spiral structure -- galaxies: rings -- galaxies:
  bars -- galaxies: dynamics -- galaxies:morphology -- galaxies:
  orbits -- manifolds -- chaos }
}
\maketitle{}

\section{Introduction}

Barred galaxies often have interesting morphological
characteristics. These include two-armed, grand design spirals and
rings, both inner and outer. Over the last five years we 
developed a theory which accounts for the formation and the
properties of these structures, starting with the dynamics of the
regions around the ends of the bar. Our main results are summarised in
four papers [Romero-G\'omez et al. 2006 (Paper I), 2007 (Paper II),
  2008; Athanassoula, Romero-G\'omez \& Masdemont 2008 (Paper III)],
while a fifth one focuses more on comparisons with observations
(Athanassoula, Romero-G\'omez, Bosma \& Masdemont 2009, 
Paper IV). In these papers, the reader will also find 
other relevant references, concerning other aspects of this problem,
either theoretical, or observational.  

As discussed in \cite{BinneyTremaine08}, barred galaxies have
five equilibrium points, often referred  
to as Lagrangian points, of which two are unstable ($L_1$ and $L_2$) and
three are stable ($L_3$, $L_4$ and $L_5$). The former are located on the
direction of the bar major axis, while the latter are located at the
centre of the galaxy and on the direction of the bar minor axis. From
the region around the $L_1$ and $L_2$ emanate the manifolds which
guide the chaotic orbits escaping this vicinity. These manifolds can
be simply thought of as tubes that guide and confine the chaotic
orbits in question. It is this confined chaos that can account for the
grand design spiral arms and the inner and outer rings.

The structures we wish to understand are the spirals and the inner and
outer rings in barred galaxies. Inner rings have roughly the size of
the bar and are elongated along it. They are noted as $r$ in the
classification of de Vaucouleurs and his collaborators \citep[][and
references therein]{ButaCO07}. Outer rings are
similar, but of larger size. They are either elongated perpendicular
to the bar ($R_1$ type), or along it ($R_2$ type). In some
cases the two types of outer rings co-exist, and the structure is then
noted $R_1R_2$. Thus, a barred galaxy with both an inner ring and an outer
ring of the first type will be noted $rR_1$. A description of this
classification, as well as more information on these structures has
been given by \cite{Buta95}. Spirals are also very common amongst
barred galaxies \citep{ElmegreenElmegreen89}. In most cases they start
off from the ends of the bar, and they always wind outwards in a
trailing sense.   

This paper is organised as follows. In Sect.~\ref{sec:models} we describe
the bar models used in our calculations. In Sect.~\ref{sec:theory} we 
present the theoretical basis of our theory. In 
Sect.~\ref{sec:twoexamples} we discuss two specific morphologies, the spirals
and the $rR_1$ rings, and the different patterns of matter circulation
that they entail. We also discuss the types of potentials that can lead
to these morphologies. 
In Sect.~\ref{sec:stable} and \ref{sec:gas}, we address two 
physical problems, namely the presence of two extra mass
concentrations centred on 
the two ends of the bar and the effect of the gas, respectively. Finally,  
we conclude in Sect.~\ref{sec:final}.

Response calculations in barred galaxy potentials were discussed in
this meeting by Patsis in his talk (these proceeedings; see also
\cite{Patsis06} and references in both), while Efthymiopoulos et
al. put up a poster discussing driving by spirals (these proceedings; see also
\cite{TsoutsisEV08} and references in both). 
  
\section{Models}
\label{sec:models}

We model the barred galaxy as the superposition of an axisymmetric component 
with a non-axisymmetric one, representing a bar rotating around the
centre of the galaxy. 

We use three different bar models to describe the bar component. The basic 
bar model is described by a Ferrers ellipsoid \citep{fer77} of density 
distribution

\begin{equation}
\rho = \left\{\begin{array}{lr}
\rho_0(1-m^2)^n & m\le 1\\
 0 & m\ge 1,
\end{array}\right.
\label{eq:Ferden}
\end{equation}

\noindent
where $m^2=x^2/a^2+y^2/b^2$. The values of $a$ and $b$
determine the shape of the bar, $a$ being the length of the semi-major
axis and $b$ being the length of the semi-minor
axis. In a frame of reference co-rotating with the bar, the major axis
is placed along the
$x$ coordinate axis. The parameter $n$ measures the degree of concentration  
of the bar, while the parameter $\rho_0$ represents the central density of the 
bar. For these models, the quadrupole moment of the bar is given by the 
expression
$$Q_m=M_b(a^2-b^2)/(5+2n), $$
where $M_b$ is the mass of the bar, equal to
$$ M_b=2^{(2n+3)}\pi ab^2 \rho_0 \Gamma(n+1)\Gamma(n+2)/\Gamma(2n+4) $$
and $\Gamma$ is the gamma function. 

This bar model is the basic model of \citet{ath92a} and it will also be our 
basic bar model. We refer to it as model A and it has essentially four free 
parameters that determine the dynamics in the bar region. The axial ratio 
$a/b$ and the quadrupole moment (or mass) of the bar $Q_m$ (or $M_b$) will 
determine the strength of the bar. The third parameter is the pattern
speed, i.e. the angular velocity of the bar. The last free parameter is the 
central density of the model ($\rho_c$). For reasons of 
continuity we will use the same numerical values for the model parameters as 
in \citet{ath92a, ath92b} and in Papers II and III. The length of the bar is 
fixed to 5 kpc. More information on these models can be found in
\citet{ath92a} and in Papers II and III.

Since we want to study the formation of spiral arms and rings and both
features are located in the outer regions of the bar or well outside
it, we consider two other  
bar potentials to avoid a well known disadvantage of Ferrers
potential, namely the fact that the non-axisymmetric component of the
force drops very steeply beyond a certain radius, particularly for
values of the density index $n$ larger than one (eq.~\ref{eq:Ferden}), so 
that the axisymmetric component dominates in the outer regions. These ad-hoc
potentials are of the form $\epsilon A(r)\cos(2\theta)$, $\epsilon$ being 
related to the strength of the bar.

The first ad-hoc bar potential we use is adapted from \citet{deh00} and has 
the form

\begin{displaymath}
\Phi(r,\theta)=-\frac{1}{2}\epsilon v_0^2\cos(2\theta)\left\{{\begin{array}{ll}
\displaystyle 2-\left(r/\alpha\right)^n, & r\le \alpha\rule[-.5cm]{0cm}{1.cm}\\
\displaystyle \left(\alpha/r\right)^n, & r\ge \alpha.\rule[-.5cm]{0cm}{1.cm}
\end{array}}\right.
\label{eq:adhoc1}
\end{displaymath}

The parameter $\alpha$ is a characteristic length scale of the bar potential 
and $v_0$ is a constant with units of velocity. The parameter
$\epsilon$ is a free  
parameter related to the bar strength. In this paper we use $\alpha$ = 5 and
$n~=~0.75$. We will refer to models with this non-axisymmetric
component as the D models. 

Finally, our third model has the bar potential:

\begin{displaymath}
\label{eq:adhoc2}
\Phi(r,\theta)=\hat{\epsilon}\sqrt{r}(r_1-r)\cos(2\theta),
\end{displaymath}

\noindent
where $r_1$ is a characteristic scale length of the bar potential, which we 
will take for the present purposes to be equal to 20 kpc. The parameter 
$\hat{\epsilon}$ is related to the bar strength. This type of model has 
already been widely used in studies of bar dynamics \citep[e.g.][]{bar67,
con80, con81}.  We will refer to models with this non-axisymmetric
component as the BW models.

Models D and BW are meant to represent and include forcings not only from
bars, but also from spirals, from oval discs and from triaxial
haloes. We have kept all these forcings bar-like, i.e. we
have not included any radial variation of the azimuthal dependence.
This was done on purpose, in
order not to bias the response towards spirals and in order to
avoid the extra degree of freedom resulting from the azimuthal
winding of the force.

Throughout this paper we use the following system of units: For the mass unit 
we take a value of $10^6 M_{\odot}$, for the length unit a value of 1 kpc and 
for the velocity unit a value of 1 km/sec. Using these values, the 
unit of the Jacobi constant will be 1 km$^2$/sec$^2$.

\section{Theoretical background}
\label{sec:theory}

\subsection{Equilibrium points and dynamics around them} 
\label{sec:eq}

We will now study the dynamics of the motion  in the equatorial plane
of the galaxy, $z=0$. We work in a frame of  
reference co-rotating with the bar, i.e. a frame in which the bar is at
rest and by convention we place it along the $x$ axis.

\begin{figure}[]
\resizebox{\hsize}{!}{
\includegraphics[clip=true,angle=-90.]{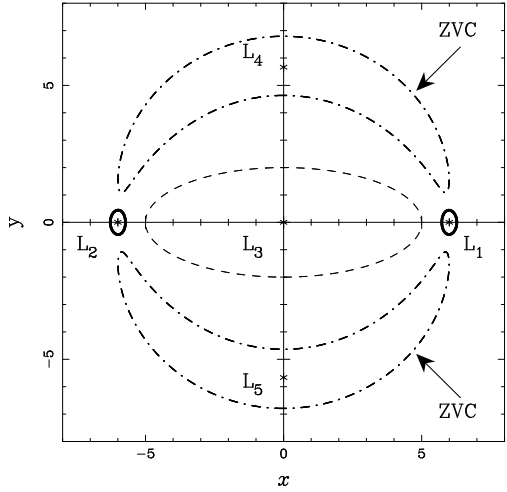}}
\caption{\footnotesize Location of the equilibrium points ($L_i$, $i =
  1,..,5$, marked with asterisks) 
in model A with $a/b$=5, $Q_m=4.5\times 10^4$, $r_L=6$ and $\rho_c=2.4\times
10^4$. The dashed line gives the outline of the bar, while 
the dot-dashed lines show the zero velocity curves for a characteristic
energy. Around $L_1$ and $L_2$ we  
plot the Lyapunov periodic orbits of this energy (thick solid lines).}
\label{fig:eq}
\end{figure}

The energy of a particle in this rotating frame is a constant of the motion
and is given by
$$E_J=\frac{1}{2}\left ( {\dot x}^2+{\dot y}^2\right )+\Phi_{\hbox{\scriptsize eff}},$$

\noindent 
where $\Phi_{\hbox{\scriptsize eff}}=\Phi-\frac{1}{2}\Omega_p^2\,(x^2+y^2)$ is
the effective potential, i.e. the potential in the rotating frame, and $\Phi$ 
is the potential \citep{BinneyTremaine08}. The curve defined by
$\Phi_{\hbox{\scriptsize eff}}=E_J$ is  
the zero velocity curve (ZVC) and divides the equatorial plane in three 
different regions, namely the forbidden region, the inner region, where the 
bar is located, and the outer region (see Fig.~\ref{fig:eq}). The
forbidden region is enclosed within the ZVC and is forbidden to
particles with energy equal to that of the ZVC, or smaller. The inner
region is delineated by the inner part of the ZVC of energy equal to
that of the $L_1$ and $L_2$ and includes the centre, while the outer
region is delineated by the outer part of this ZVC and includes the
outermost parts of the galaxy.

The system has five equilibrium points located on the equatorial plane where

$$\frac{\partial \Phi_{\hbox{\scriptsize eff}}}{\partial x}=
\frac{\partial \Phi_{\hbox{\scriptsize eff}}}{\partial y}= 0.$$

\noindent
$L_1$ and $L_2$ lie on the direction of the bar major axis and they are saddle
points, i.e. linearly unstable. The other three are stable. $L_3$ is
at the origin of coordinates and it is a minimum of the effective
potential, while $L_4$ and $L_5$, located along the direction of the
bar minor axis  
(see Fig.~\ref{fig:eq}), are maxima. We define $r_L$ as the distance
from the origin to $L_1 (L_2)$. 

Around each of the equilibrium points, there exists a family of periodic 
orbits. For example, around $L_3$ we have the $x_1$ family of stable
periodic orbits \citep{con80},
which is known to be responsible for the bar structure 
\citep{ath83}. In the linear approximation, the motion around
$L_1$ and $L_2$ consists of the  
superposition of an exponential part and an oscillation:

\begin{displaymath}
\label{eq:sol-lin}
\left\{
{
\begin{array}{l}
x(t)  =  X_1 e^{\lambda t} + X_2 e^{-\lambda t}
        +X_3\cos(\omega t + \phi), \\
y(t)  =  A_1X_1 e^{\lambda t} - A_1X_2 e^{-\lambda t} \\
        ~~~~~~~~~~~+A_2X_3\sin(\omega t +\phi),
\end{array}}
\right.
\end{displaymath}

\noindent
where $\pm \lambda$, $\pm \omega i$ are the eigenvalues of the differential
matrix at $L_1 (L_2)$, and $X_1, X_2, X_3, A_1$, and $A_2$ are constants. For 
each energy level there exists a unique periodic orbit around $L_1$, or $L_2$, 
called Lyapunov orbit \citep{Lyapunov49}. To obtain them, we
take initial conditions from Eq.~\ref{eq:sol-lin} with $X_1=X_2=0$:

\begin{displaymath}
{
\begin{array}{ll}
{\bf x}(t)=& (x,y,\dot{x},\dot{y})=\\
&(X_3 \cos(\omega t+\phi), \\
&\, A_2X_3 \sin(\omega t+\phi), \\
&\,-X_3\omega\sin(\omega t+\phi), \\
&\, A_2X_3 \omega\cos(\omega t+\phi)).
\end{array}}
\end{displaymath}

These orbits can of course be calculated numerically directly from the
equations of motion in the adopted potential and reference frame,
without the use of the linear approximation. Examples are shown in
Fig.~\ref{fig:eq}.
They are unstable, so they cannot trap particles like a stable family. 
However, there are other objects emanating from the periodic orbits that can 
trap particles. They are called the invariant manifolds and are
associated to the  
periodic orbits. From each periodic orbit emanate four branches of asymptotic 
orbits, two of them being stable and the other two unstable. The 
characterisation of stable and unstable in this definition is in the sense 
that the stable invariant manifold is a set of orbits that approach 
asymptotically the periodic orbit, while the unstable invariant manifold is a 
set of asymptotic orbits that depart asymptotically from the periodic
orbit. That is, there are four favourite directions along which
material can approach or escape from the vicinity of the corotation
region and they are given by the stable and the unstable 
manifolds, respectively. 

In the linear case, we can obtain an initial
condition for 
the stable invariant manifold considering $X_1=0$ and $X_2\neq 0$
in Eq.~\ref{eq:sol-lin}. The exponential term proportional to $X_2$ vanishes 
when time tends to infinity and the trajectory tends to the periodic orbit.
Analogously, if we consider initial conditions with $X_1\neq 0$ and $X_2= 0$, 
we will obtain the unstable invariant manifold.

Thus, the manifolds can be thought of as tubes which guide chaotic
orbits escaping the vicinity of $L_1$ and $L_2$. Therefore, according to
this theory, spiral arms will be due to confined chaos.
  
\subsection{Transfer of matter. Homoclinic and heteroclinic orbits}
\label{sec:homohetero}

\begin{figure*}[]
\resizebox{\hsize}{!}{
\includegraphics[clip=true,angle=-90.]{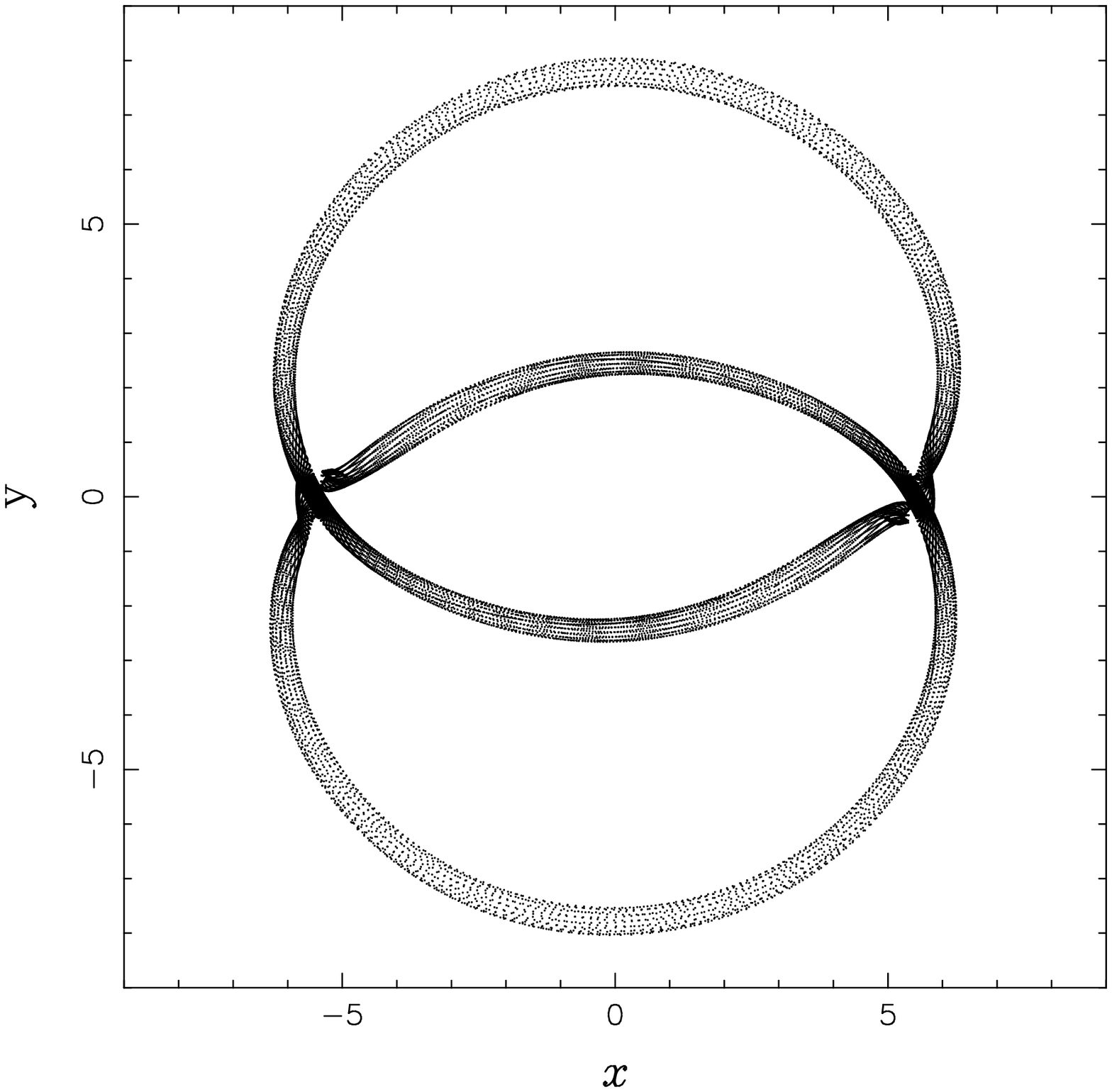}
\includegraphics[clip=true,angle=-90.]{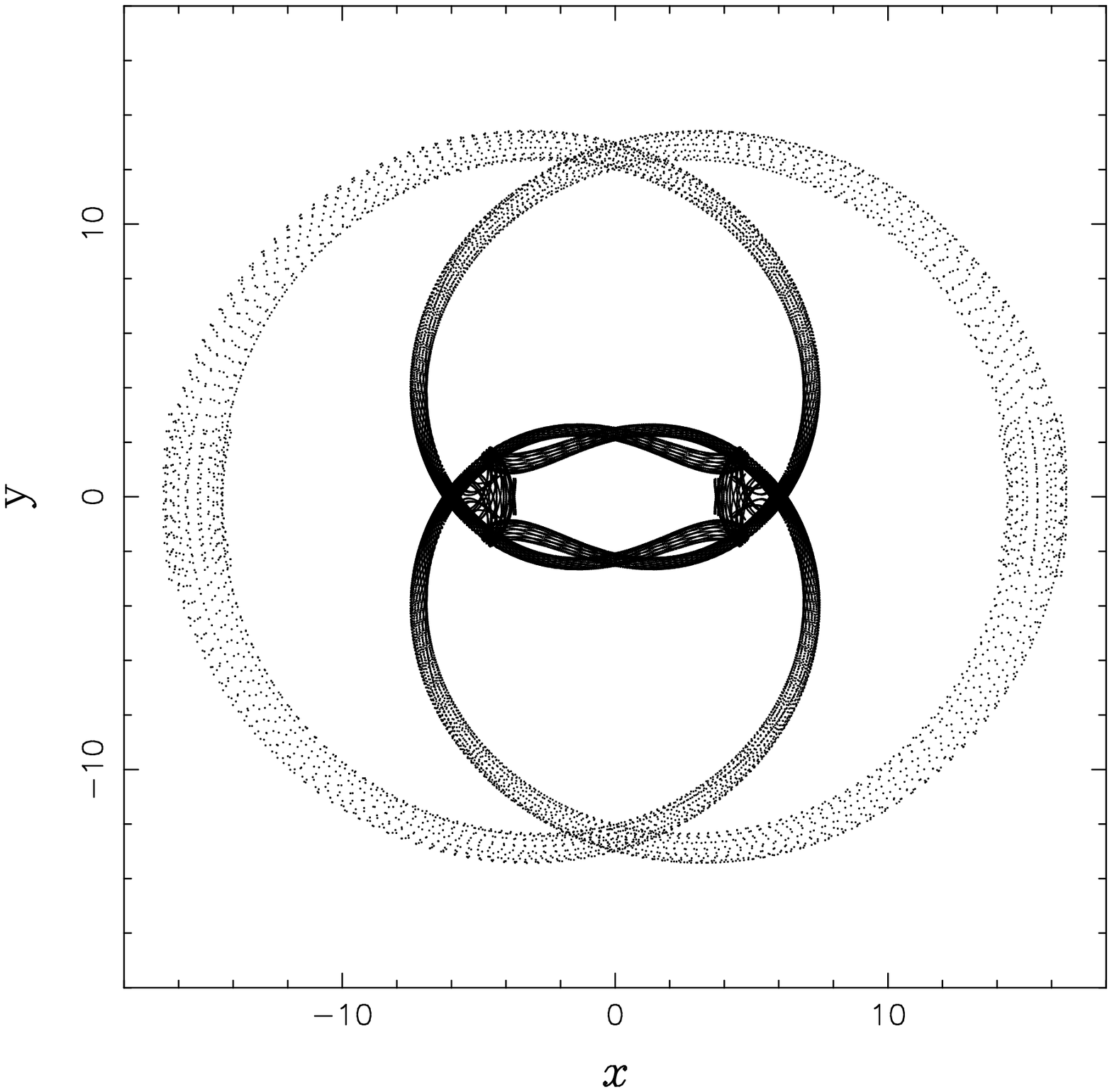}
\includegraphics[clip=true,angle=-90.]{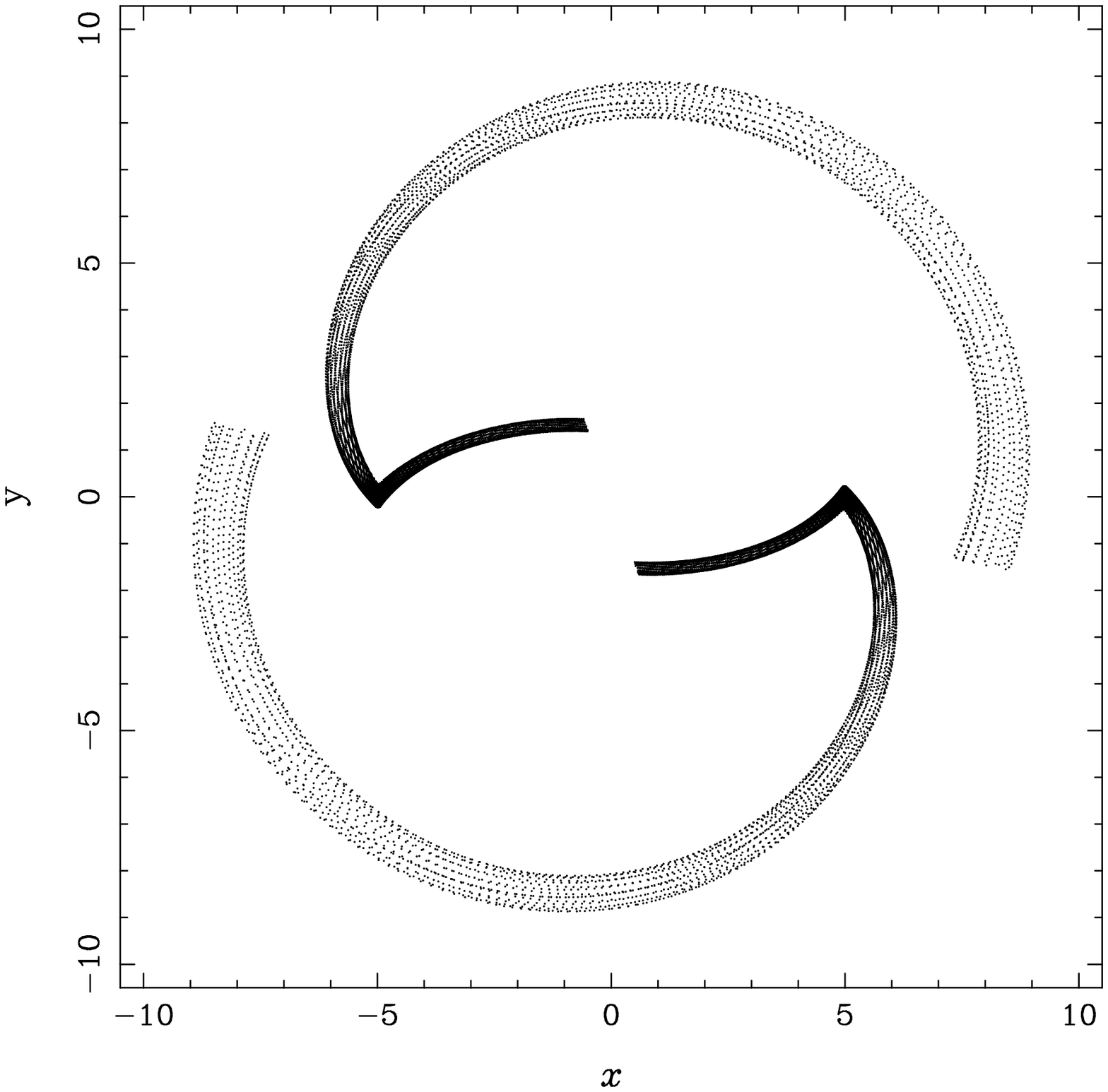}
\includegraphics[clip=true,angle=-90.]{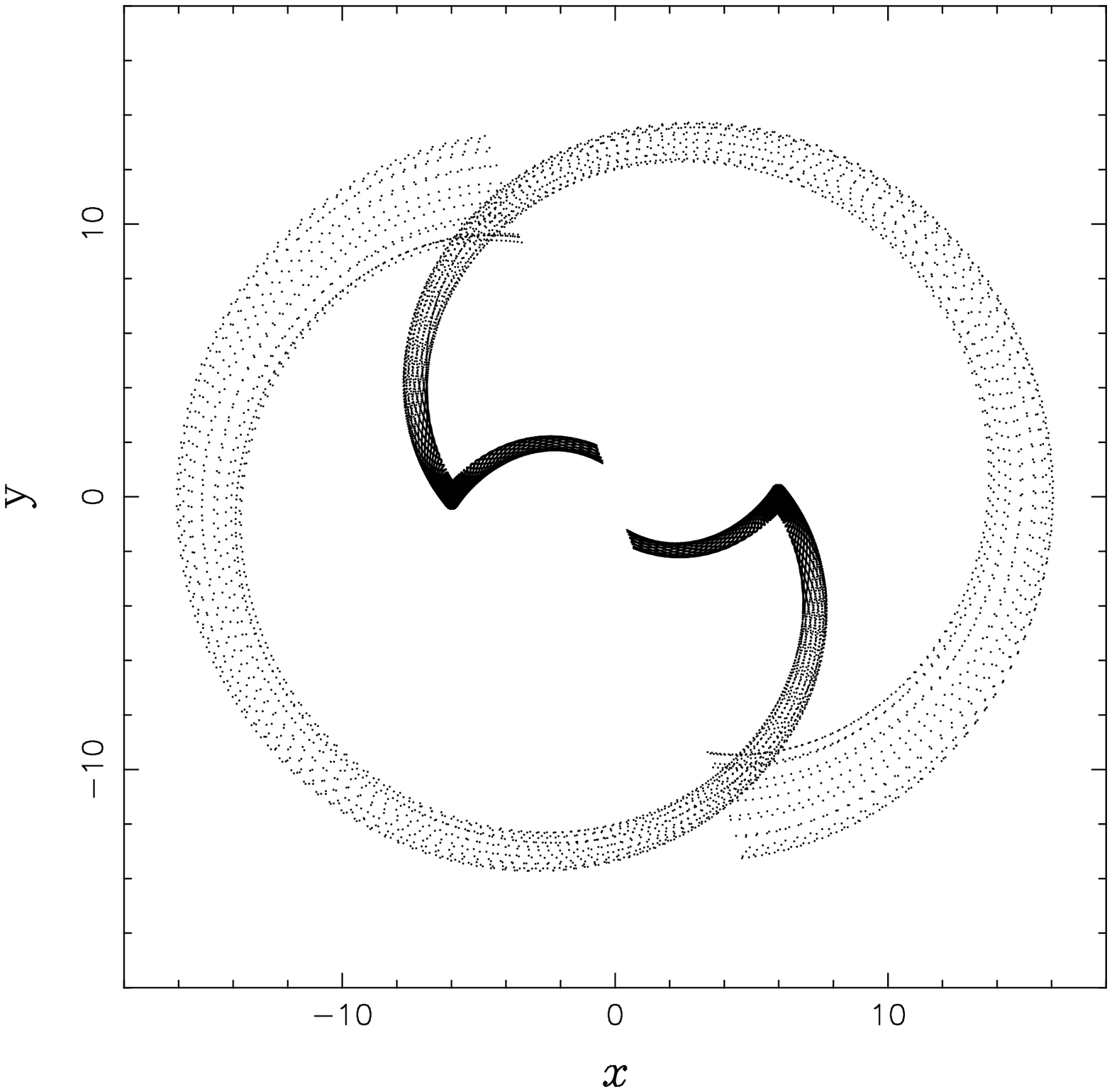}
}
\caption{\footnotesize Four different morphologies.
{\sl First panel}: Model with heteroclinic orbits having
an $rR_1$ morphology. {\sl Second panel}: Model with homoclinic orbits
having an $R_1R_2$ ring  
morphology. {\sl Third panel}: Model with no heteroclinic or
homoclinic orbits having two spiral arms. {\sl Fourth panel}: Model with 
no heteroclinic or homoclinic orbits having an $R_2$ ring.}
\label{fig:mor}
\end{figure*}

Invariant manifolds are not restricted to the neighbourhood of $L_1 (L_2)$, but
they extend well beyond the region around the equilibrium points. Two of the
branches, one stable and one unstable, are located in the inner region, while 
the other two are located in the outer region, so that the invariant manifolds
connect the two regions and the periodic orbits act as a gateway.
In Fig.~\ref{fig:mor}, we can see four examples of the morphologies 
obtained with these invariant manifolds. 

The stable (respectively, unstable) manifolds of a periodic orbit cannot 
intersect in the phase space. However, the stable and the unstable 
branches can intersect each other. If these stable and unstable
branches are associated to two different periodic orbits, one around
the $L_1$ and the other around the $L_2$, we obtain heteroclinic
orbits, i.e., orbits that connect one of the ends of the bar 
with the opposite one. Due to the symmetry of the system, the 
particles following these orbits will outline a morphology similar to
that of an $R_1$ ring (left panel of Fig. \ref{fig:mor}).

If, on the other hand, the stable and unstable branches are 
associated to the same Lyapunov periodic orbit, the intersection gives
rise to what is known as 
homoclinic orbits. That is, asymptotic orbits that connect one of the bar 
ends with itself. Considering the symmetry of the system, we will have 
particles outlining trajectories reminiscent to that of $R_1R_2$ rings, i.e. 
the particles will form two outer rings, one with major axis perpendicular to
the bar major axis ($R_1$ ring) and one with major axis parallel to it ($R_2$
ring), as shown in the second panel of Fig.~\ref{fig:mor}. 

If the 
stable and unstable manifolds emanating from one of the Lyapunov
orbits do not return either to that orbit, or to the corresponding one
around the Lagrangian point at the opposite side of the bar, but
unwind outwards, they form a spiral shape (third panel of
Fig. \ref{fig:mor}). Thus, such morphologies can give rise to two
trailing grand design spirals. We will call orbits following such
manifolds escaping, because they can escape the vicinity of the bar and
reach the outer parts of the galaxy.  

Finally, if the outer branches of the
unstable invariant manifolds emanate with an appropriate pitch angle, they 
will intersect in the configuration space forming an outer ring whose major
axis is parallel to the bar major axis, i.e. an $R_2$ ring (right
panel of Fig. \ref{fig:mor}).

\section{Two specific examples}
\label{sec:twoexamples}

As described in the previous section, manifolds can reproduce the right
morphologies for both spirals and rings (inner, $R_1$, $R_2$,
$R_1R_2$). In this section we will consider in more detail two
possible morphologies, $rR_1$ rings and spirals.

An $rR_1$ morphology necessitates heteroclinic orbits. In this case,
the stable and unstable  
manifolds overlap in the configuration space. The inner branches form
an inner ring, elongated along the bar major axis. The outer branches
form an $8$-like or
$\Theta$-like shape, i.e. an $R_1$ ring morphology. In particular,
this is elongated perpendicular to the bar major axis (i.e. its major
axis is perpendicular to that of the bar) and it also produces
characteristic dimples on either side of the bar, at or near the ends
of the inner ring. Such dimples have also been found, in the same
location, in observed galaxies \citep{ButaCrocker91}.

\begin{figure*}[]
\resizebox{\hsize}{!}{
\includegraphics[clip=true,angle=-90.]{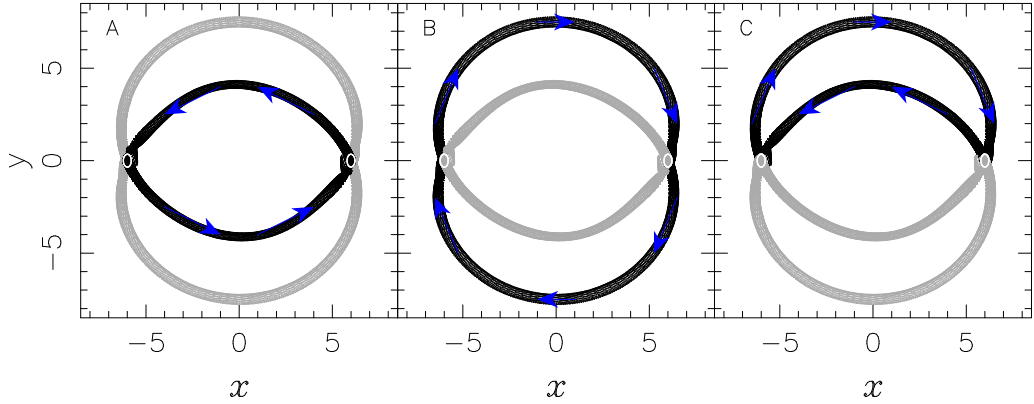}}
\caption{\footnotesize The four possible circulation patterns for particles 
on orbits guided by the manifolds. Any given such particle can stay on
the inner branches of the manifold, i.e. can trace the inner ring
(left panel), or it can stay on the outer branches of the manifold,
i.e. can trace the outer ring  (middle panel). Alternatively, it can
trace a path that includes one inner and one outer branch. Two such
paths are possible. One is shown in the right panel. The other, i.e. the fourth 
alternative, is not plotted here since it is similar to the third one, except 
that it is symmetric with respect to the bar major axis. After a
particle has completed a full circulation path, it can either repeat
the same path, or take any of the other three ones. The manifolds
coming into play in each case are plotted in black in the
corresponding panel, and the remaining ones in light grey. In white,
we plot the Lyapunov periodic orbits of the corresponding
energy level. The arrows show the direction of the motion.}
\label{fig:circring}
\end{figure*}

This type of manifold morphology leads to interesting circulation patterns.
Particles can follow essentially four different paths, 
shown in Fig.~\ref{fig:circring}. One possibility is that they 
circulate along the inner branches of the invariant manifolds forming the 
inner ring (left panel of Fig.~\ref{fig:circring}). This circulation
is anti-clockwise. The second possibility 
is that they follow the outer branches of the manifolds, outlining the outer 
ring (middle panel of Fig.~\ref{fig:circring}). This circulation is
clock-wise. Finally, they can have 
a mixed trajectory, i.e. follow both inner and outer branches. Thus,
particles starting from $L_1$ and following the inner branch of the
unstable manifold approach the neighbourhood of $L_2$, move out of
the bar region and
follow the outer branch of the unstable manifold. They then reach a maximum
distance from the centre and come back to the neighbourhood of $L_1$ 
(right panel of Fig.~\ref{fig:circring}). In this way the circulation
pattern is completed. A fourth circulation pattern (not shown in
Fig.~\ref{fig:circring}) has an identical shape but is reflected with
respect to the bar major axis. In both the third and the fourth cases
the circulation is clockwise with respect to their
`centres', i.e. with respect to the $L_3$ and $L_4$ Lagrangian points,
respectively.  

There are, thus, four circulation paths in total, one using
only the inner branches, one using only the outer branches and two
mixed ones, using both inner and outer branches. Once one such
pattern is completed, a particle can
either repeat it or take any of the other three circulation patterns.
Thus, material from inside corotation can move outside it and
vice-versa. However, the maximum radius such material can reach is
bound by the maximum distance of the path from the centre, located on
the direction of the bar minor axis. 

As a second example, let us discuss a spiral morphology. Material,
initially on an inner stable manifold branch, can, via the
neighbourhood of one of the unstable Lagrangian points, move to an
outer unstable branch. If this is of the escaping type (see previous
section), we obtain a morphology
similar to that of the grand design spiral arms of a barred galaxy.

\begin{figure}[]
\resizebox{\hsize}{!}{
\includegraphics[clip=true,angle=-90.]{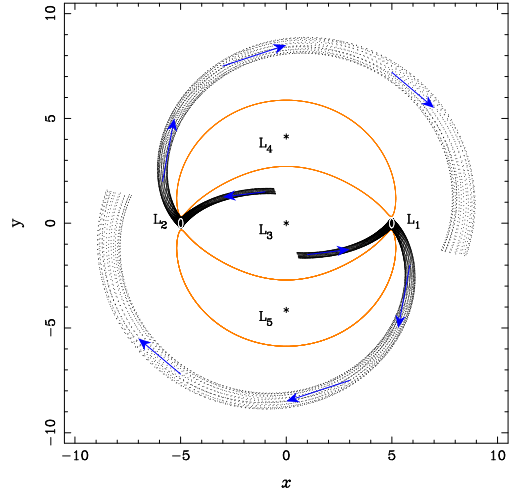}}
\caption{\footnotesize Circulation pattern for particles in models with
  spiral arms. The particle will move along the inner stable manifold branch
  and reaching the vicinity of one of the two unstable Lagrangian
  points will continue outwards along the unstable branch of the
  manifold. The direction of the motion is given by the arrows. We also
  plot the ZVC of the same energy as the manifolds (solid line) and give
  the position of the Lagrangian points with asterisks.}
\label{fig:circsp}
\end{figure}

The circulation pattern in this case is much simpler
(Fig.~\ref{fig:circsp}). Matter from the 
inner region (more specifically from the outer regions of the bar or
its immediate vicinity) moves along the inner manifolds towards one of
the two unstable Lagrangian points and from there onto the
corresponding outer branch of the unstable manifold. It can thus
escape the inner region and reach outer parts of the disc. 

Of course,
an inwards going route could also be possible, at least in
principle. Then matter from the outer parts of the disc could move
inwards along a stable outer manifold branch to the $L_1$ or the $L_2$
and from there into the inner region. However, as will be discussed
further in Sect.~\ref{sec:final}, the existence of a given manifold
does not necessarily imply the existence of the corresponding
structure in a galaxy, or in an $N$-body simulation. For this, the
manifolds have to confine a sufficient amount of matter. This is similar
to periodic orbits, where the existence of a stable family 
will not imply the formation of any structure if it does not
trap any regular orbits around it. Thus the route bringing matter
inwards, although in principle possible, would imply the existence of
leading two-armed grand design spirals in barred potentials and has
not yet been observed either in simulations or in real galaxies. 
Whether and how matter gets trapped by periodic orbits, or by
manifolds depends on the formation history of the galaxy and is well
beyond the scope of this paper. In Paper IV, however, we examine a few
specific cases. 
 
The two different morphological types discussed above, namely $rR_1$
rings and two-armed grand design spirals, do not come in the same
potentials. The $rR_1$ morphology comes in models with bar potentials
which are not too strong in the corotation region and immediately
beyond it. In Paper III we gave, for the types of models of
Sect.~\ref{sec:models}, upper limits of the bar strength beyond which
this morphology will not 
occur. On the other hand, spirals form in models with stronger bar
or appropriate spiral forcings. We also showed that it is the strength of the
non-axisymmetric potential at and immediately beyond corotation that
are the best indicators of the morphological type. For this we use the
quantity 

\begin{equation}
Q_t (r) = (\partial \Phi (r, \theta) /\partial \theta)_{max}/(r\partial
\Phi_0/\partial r),
\label{eq:Qt}
\end{equation}

\noindent
where $\Phi$ is the potential, $\Phi_0$ is its axisymmetric part and the
maximum in the numerator is calculated over all values of the azimuthal angle 
$\theta$. We calculate $Q_t (r)$ at the radius of $L_1$, i.e. at
$r = r_{L1}$ and denote it by $Q_{t,L_1}$. As shown in Paper III,
$R_1$ morphologies form in models with relatively low
values of $Q_{t,L_1}$, while higher values of this quantity
give rise to other morphologies.

\section{Cases with stable $L_1$ and $L_2$ Lagrangian points}
\label{sec:stable}

\subsection{Achieving stability}
\label{subsec:stability}

The morphologies and the characteristics of the invariant manifolds described
so far assume that the $L_1$ and $L_2$ equilibrium points are unstable
and with linear stability of the type
``saddle$\times$centre$\times$centre''. We will hereafter 
refer to this case as the standard case. To make our theory more general, we 
also examine under what conditions the $L_1$ and $L_2$ can become 
stable and what consequences this will have on the galaxy morphology
(Paper III).

It is possible to stabilise the equilibrium points by adding a concentration
of matter around them. We tested this by adding two identical, small 
Kuz'min/Toomre discs \citep{kuz56,too63}, one centred on each of the $L_1$ and
$L_2$. By increasing their mass, or decreasing their scale-length,  
the equilibrium point $L_1$ bifurcates,
becoming stable, while two new unstable points appear, both on the
direction of the bar major axis. They are located one on either side
of $L_1$, and are called $L_1^i$ and $L_1^o$, the
subscript, $i$ or $o$, denoting whether the corresponding new
Lagrangian point is located inside or outside the $L_1$. 
Analogously, $L_2$ also becomes stable and two more unstable equilibrium 
points, $L_2^i$ and $L_2^o$, appear on either side of it. The mass
of the added concentration at which this is attained is a decreasing
function of its scale-length. Similarly, the concentration
scale-length at which this is attained is 
an increasing function of its mass. 

\begin{figure}[]
\resizebox{\hsize}{!}{
\includegraphics[clip=true,angle=-90.]{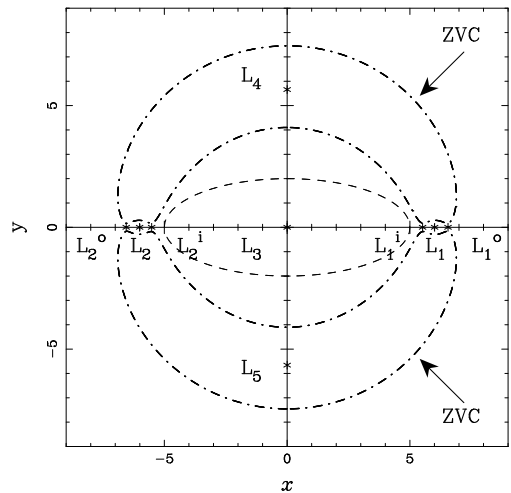}}
\caption{\footnotesize Location of the equilibrium points (marked with
  an asterisk)
in an A model with $a/b=2.5$, $r_L=6$, $Q_m=4.5\times 10^4$ and 
$\rho_c=2.4\times 10^4$, to which have been added two identical
Kuz'min/Toomre discs centred at the $L_1$ and $L_2$ Lagrangian points.
The scale length and the mass of these discs are 0.6 and $0.025M_b$,
respectively. The dashed line  
marks the position of the bar, while the dot-dashed lines show the zero 
velocity curves. }
\label{fig:eq2}
\end{figure}

This new configuration is shown in 
Fig.~\ref{fig:eq2}, where we plot the ZVC for a given energy, the position of 
the equilibrium points and the outline of the bar, for model A with $a/b=2.5$,
$r_L=6$, $Q_m=4.5\times 10^4$, and $\rho_c=2.4\times 10^4$. The scale-length
and the mass of the two identical Kuz'min/Toomre discs are 0.6 and $0.025M_b$,
respectively. The linear stability of the four new equilibrium points is of the 
type ``saddle$\times$centre$\times$centre'', so we can compute the family of 
unstable periodic orbits and the manifolds associated to them, as we
did for those of $L_1$ and $L_2$ in the standard case.

\subsection{Global morphology}
\label{subsec:glmorphostab}

\begin{figure*}[]
\resizebox{\hsize}{!}{
\includegraphics[clip=true,angle=-90.]{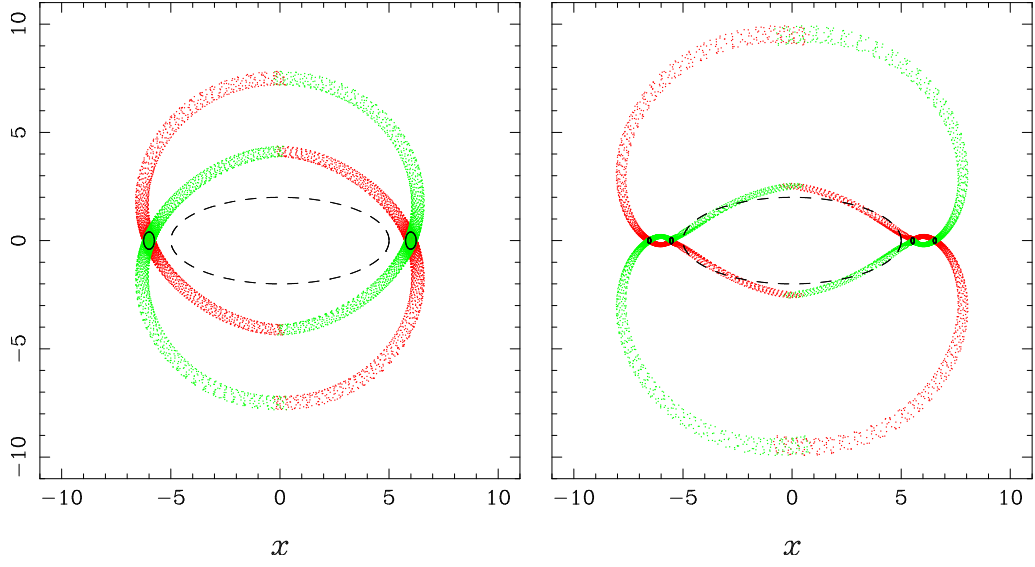}}
\caption{\footnotesize Comparison of the manifold morphology in the standard
model with 5 equilibrium points (left panel, equilibrium points as in
Fig.~\ref{fig:eq}) and with 9 equilibrium points
(right panel, equilibrium points as in Fig.~\ref{fig:eq2}). The bar
outline is given by a dashed line and the corresponding Lyapunov
orbits by a black solid line. }
\label{fig:stable}
\end{figure*}

In Fig.~\ref{fig:stable} we compare the morphology obtained with the standard
case to that with the same bar model plus the two mass concentrations
around the $L_1$ and $L_2$ in the form of two small Kuz'min/Toomre discs. The
two morphologies are essentially of the same type, since in both cases
we obtain an $rR_1$ structure. However, the sizes and axial ratios of the
two rings change drastically when the two mass concentrations are added
and the number of equilibrium points increases to 9. The inner ring becomes smaller and
is located very near the outline of the bar. Its axial ratio
approaches that of the bar, while in the `standard' case it has a
smaller ellipticity. The major axis of the outer ring becomes much larger, and
its axial ratio changes accordingly. If no material circulates between
$L_1^i$ and $L_1^o$ and between $L_2^i$ and $L_2^o$, the inner
and outer rings will not join, i.e. the outer ring will be detached.
Such detached rings are often observed and this may be the way, or one
of the ways, to explain their formation. 

\subsection{Specific morphology around the $L_1$ and $L_2$}
\label{subsec:locmorphostab}

\begin{figure}[]
\resizebox{\hsize}{!}{
\includegraphics[clip=true,angle=-90.]{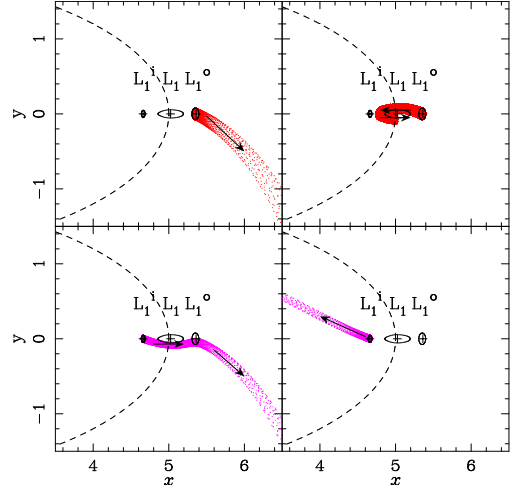}}
\caption{\footnotesize Morphology of the unstable manifold branches in the
  vicinity of a stable $L_1$ Lagrangian point (see text). The two
  upper panels show the manifolds linked to $L_1^o$ and the two lower
  ones those linked to $L_1^i$. The arrows show the direction of the
  motion around the manifolds and the dashed line shows the outline of
  the bar. Three periodic
  orbits -- one around each of the $L_1$, $L_1^i$ and $L_1^o$ points --  
  are given with solid lines. }
\label{fig:unstlocal}
\end{figure}

\begin{figure}[]
\resizebox{\hsize}{!}{
\includegraphics[clip=true,angle=-90.]{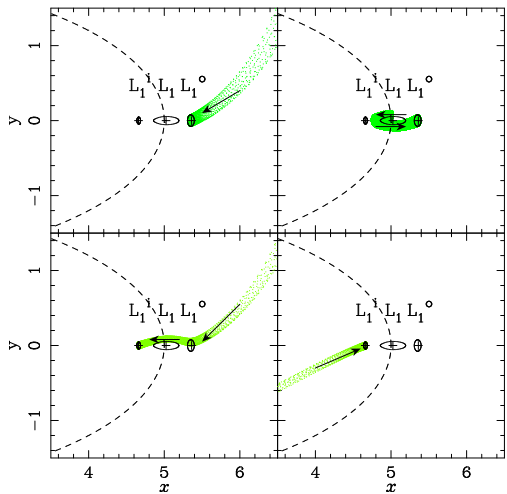}}
\caption{\footnotesize 
  Same as Fig.~\ref{fig:unstlocal}, but for the stable branches of the
  manifolds. }
\label{fig:stlocal}
\end{figure}

The morphology of the unstable and stable manifolds in the
immediate neighbourhood of the $L_1$ Lagrangian point is shown in
Figs.~\ref{fig:unstlocal} and \ref{fig:stlocal}, respectively, for a
model which has a spiral morphology. This has several similarities, but
also several differences from the model shown in Figs. 14 and 15 of 
Paper III, which is for a less strong bar with manifolds of $rR_1$
morphology. 

The Lyapunov periodic orbits around $L_1$ are elongated in the same
direction as the bar, while the periodic orbits around $L_1^i$ and
$L_1^o$ are elongated perpendicular to it. It is the one around $L_1$
that is the largest and the one around $L_1^i$ that is the smallest,
the third one being of intermediate size. 

There are now in total four unstable branches (Fig.~\ref{fig:unstlocal}). 
The outer branch of the $L_1^o$ manifold, as well as the inner branch
of the $L_1^i$ have a very simple morphology, while the two other
branches have a more complicated and interesting morphology. The inner
branch of the $L_1^o$ emanates from the vicinity of $L_1^o$ and
circumventing the $L_1$ from the above extends towards
$L_1^i$, but before reaching it turns back towards $L_1^o$, now 
circumventing the $L_1$ from below. The unstable outer branch of
$L_1^i$ goes towards $L_1^o$ circumventing $L_1$ from below and then
moves outwards beyond the corotation region. The morphology of the stable
branches is similar and is shown in Fig.~\ref{fig:stlocal}.

\section{Ersatz for gas}
\label{sec:gas}

Observations show that gas is intimately linked to both spirals and
rings. For example, galaxies with smooth, very gas poor spiral arms,
known as anemic, are a rarity \citep{vdBergh76, ElmegreenEFEPGI02}.
Furthermore, the formation of rings and spirals from gas has been
witnessed in a number of simulations \citep[e.g.][]{sch81,
  com85, ath92b, LindbladLA96, WadaKoda01, LinYuanButa08,
  RautiainenSL08}. We should thus compare  
the dynamics of the gas with that of the manifolds presented here. 
Gas, however, has different equations of motion from stars, so we need
to modify our calculations accordingly. 

\citet{sch79,sch81,sch84,sch85a} 
uses sticky particles to simulate the gas and models collisions in a 
particularly straightforward way, so we can introduce a similar procedure 
also in our calculations. In Schwarz's simulations, particles represent 
gaseous clouds which lose a certain fraction of their kinetic energy when 
they collide. In practise, they lose a certain fraction $f$ of their 
velocity \citep{sch81}, or only of the component of this velocity that
is along the line joining the two particles \citep{sch84,sch85a}. 
Thus, $v_2=\pm (1-f)v_1$, where $+$ is for the tranverse velocity
component, $-$ is for the component along the line joining the two
particles, and $v_1$ and $v_2$ are the velocities 
before and after the collision, respectively. The values of $f$ in
Schwarz's simulations range between 0.8 and 1. 

We introduce collisions in our simulations in a similar way, thus
obtaining an ersatz for gas, as described in more detail in Papers III
and IV. We also use the same models as Schwarz in order to be able to
make comparisons. Here we present results for his standard model with 
$a$ = 2.6, $q$ = 0.1, $\Omega_p$ = 0.27 and $f$ = 0.8, for which he
gives sufficient results and information in his papers to allow comparisons,
and which has an $rR_1$ manifold morphology. In this model we
calculated a number of orbits of the outer branch of the unstable
manifold and drew random numbers to find the locations along 
its trajectory where each particle will undergo a collision.  
In the particular example illustrated here we have three
collisions per half bar rotation. We tried, however, different numbers
of collisions, around the values used by Schwarz, and found
qualitatively similar results. 

\begin{figure}[]
\resizebox{\hsize}{!}{
\includegraphics[clip=true,angle=-90.]{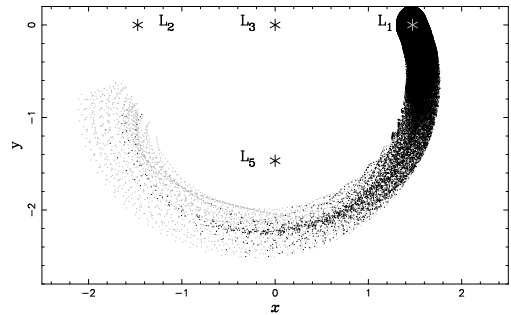}}
\caption{\footnotesize Comparison of the outer ring loci as calculated with 
(black) and without (grey) collisions. In this example $f$ = 0.8. The
  positions of the Lagrangian 
  points are marked by asterisks. See text for a description of the
calculations. }
\label{fig:colnocol}
\end{figure}

To determine the result of a collision, we take a small box around 
the collision position and calculate the average velocity of all orbits in 
that box. We then decrease the velocity of the particle relative to
that mean by a factor $\pm (1 - f)$. Fig.~\ref{fig:colnocol} shows a
comparison of the particle positions when  
they are considered as stellar (i.e. without collisions; grey) and as gaseous 
(with collisions; black). For this comparison we used $f$ = 0.8, which leads to 
a mean energy dissipation per particle of 1.25 $\times 10^{-3}$, in good 
agreement with the numbers given by \cite{sch81}. Other numerical values give
qualitatively similar results. Fig.~\ref{fig:colnocol} shows that there is in 
general good agreement between the loci of the gaseous and of the stellar
arm. Furthermore, the gaseous arm is more concentrated than the 
stellar arm, more so for a larger number of collisions per revolution. 

Physically the above results can be understood as follows. Sticky
particles (i.e. gaseous particles) follow the same orbits as
the stellar ones, except at the time of the collisions. This ensures a
general similarity. Due to the collisions, however, the gaseous particles lose
part of their kinetic energy and their velocity approaches that of the
mean. In Paper I, we compared the loci of manifolds for different
energies and found that the ones with the smaller energies lie in
configuration space {\it within} the ones with the higher
energies. This means that the corresponding spiral arms, or rings
will be thinner for the lower energies. Thus, when the particles lose
energy they will fall onto an orbit nearer to the mean and the arms
will become thinner. This is exactly what is found with the
calculations leading to Fig.~\ref{fig:colnocol}. Thus one can think of
the lowest energy manifolds (i.e. the ones having the energy of $L_1$
and $L_2$) as an attractor, to which the
gaseous trajectories will tend to because of the dissipation due to
the collisions. Thus the gaseous arms should be thinner that the
stellar ones, and this is indeed observed in real galaxies. This is
discussed more extensively in Papers III and IV, where explicit comparisons are
made.

\section{Final remarks}
\label{sec:final}

In this short review and, particularly, in the five papers on which it
is based \citep[][and Athanassoula et al. 2009]{rom06, rom07, rom08, ath08} we presented
building blocks that can explain the formation of rings (both inner
and outer) and spirals in barred galaxies. The invariant manifolds associated with
the Lagrangian points $L_1$ and $L_2$ guide chaotic orbits which have
the right shape for inner and outer rings and spirals and thus can be
their building blocks.  
In particular, here we discussed morphologies of $rR_1$ type and
of spiral type. We also discussed the case when the $L_1$ and $L_2$
are stable, and four other unstable points bifurcate from them. The
whole structure then has nine Lagrangian points, five stable and four
unstable. We also introduced an ersatz for gas and showed that the
gaseous arms and the stellar arms should have roughly the same shape
and that the former should be thinner than the latter. 

Even though dynamically sound and very
appealing, this theory will be useful for barred galaxies only if the
manifolds and the orbits they guide create structures whose properties
are in good agreement with observations. This point will be fully
addressed in paper IV. Here let us just underline a couple of noteworthy
points.

The first is that the existence of the manifolds, even if they have
the right shape and properties, does not necessarily imply that the
corresponding 
structures will be present in the galaxy. Indeed, the manifolds are only the
building blocks, and the structure will be present only if some mass
elements (e.g. stars) follow them. This is the same as for periodic
orbits, which need to trap material around them to form the corresponding
structure. Whether such material will exist or not depends on the
formation and evolution of the galaxy (see Paper IV).

The second is that this is not the only theory attempting to explain
spirals and rings, although most theories attempt either the one or
the other. Our theory relies on the existence of a
bar; in other cases a companion is necessary. The fact that a theory
is correct and reproduces well many of the main observational data
does not necessarily imply that all other theories are wrong, or
irrelevant. Nor is it necessary in order to establish a theory to show
that all others are wrong. Indeed spirals in different galaxies may
have different origins, and even in the same galaxy more than one
theory can be at work.

\begin{acknowledgements}

EA thanks Scott Tremaine for a stimulating discussion on the manifold
properties and E. M. Corsini and V. Debattista for inviting her to this
stimulating meeting. We also thank Albert Bosma and Ron Buta for very useful
discussions and email exchanges on the properties of observed
rings. This work was partly supported by grant ANR-06-BLAN-0172,  
by the Spanish MCyT-FEDER Grant MTM2006-00478, by a ``Becario
MAE-AECI'' to MRG, and by an ECOS/ANUIES grant M04U01.  

\end{acknowledgements}

\bibliographystyle{aa}

\begin{thebibliography}{}

\bibitem[Athanassoula(1992a)]{ath92a} Athanassoula, E.\ 1992, \mnras,
259, 328 

\bibitem[Athanassoula(1992b)]{ath92b} Athanassoula, E.\ 1992, \mnras,
259, 345 

\bibitem[Athanassoula et al.(1983)]{ath83} Athanassoula, E. et al.\ 
  1983, \aap, 127, 349

\bibitem[Athanassoula et al.(2008)]{ath08} Athanassoula,
  E. Romero-G\'omez, M, Mas\-demont, J. J. 2008,  
\mnras, in press and astro-ph/0811.4056 (Paper III) 

\bibitem[Barbanis \& Woltjer(1967)]{bar67} Barbanis, B., Woltjer, L.\ 1967, 
\apj, 150, 461

\bibitem[Buta(1995)]{Buta95} Buta, R. 1995, \apjs, 96, 39

\bibitem[Buta, Corwin \& Odewahn (2007)]{ButaCO07} Buta, R. J., Corwin Jr.,
  R. G., Odewahn, S. C. 2007, The de Vaucouleurs Atlas of Galaxies,
  Cambridge University Press, New York

\bibitem[Buta \& Crocker(1991)]{ButaCrocker91} Buta, R., Crocker,
  D.A. 1991, \aj, 102, 1715 

\bibitem[Binney \& Tremaine(2008)]{BinneyTremaine08} Binney, J., Tremaine, S. 2008, 
Galactic Dynamics, Second Edition, Princeton Univ. Press, Princeton

\bibitem[Combes \& Gerin(1985)]{com85} Combes, F., Gerin, M.\ 1985, 
\aap, 150, 327

\bibitem[Contopoulos \& Papayannopoulos(1980)]{con80} Contopoulos, G., 
Papayannopoulos, T.\ 1980, \aap, 92, 33

\bibitem[Contopoulos(1981)]{con81} Contopoulos, G.\ 1981, \aap, 102, 265

\bibitem[Dehnen(2000)]{deh00} Dehnen, W.\ 2000, \aj,
119, 800

\bibitem[Elmegreen \& Elmegreen(1989)]{ElmegreenElmegreen89} Elmegreen, B.G.,
  Elmegreen, D.M. 1989, \apj, 342, 677

\bibitem[Elmegreen et al.(2002)]{ElmegreenEFEPGI02}
  Elmegreen, D.M. et al. 
2002, AJ, 124, 777

\bibitem[Ferrers(1877)]{fer77} Ferrers, N.~M.\ 1877, Quart. J. Pure Appl. Math.,
14, 1 

\bibitem[Kaufmann \& Contopoulos(1996)]{KaufmannContopoulos96} Kaufmann, D.E.,
  Contopoulos, G. 1996, \aap, 309, 381

\bibitem[Kuz'min(1956)]{kuz56} Kuz'min, G.\ 1956, Astron. Zh., 33, 27

\bibitem[Lin, Yuan \& Buta(2008)]{LinYuanButa08}Lin, L-H., Yuan, C.,
  Buta, R. 2008, \apj, 684, 1048

\bibitem[Lindblad, Lindblad \&
  Atha\-nassoula(1996)]{LindbladLA96} Lindblad, P. A. B., Lindblad, P. O.,
  Athanassoula, E. 1996, \aap, 313, 65

 \bibitem[Lyapunov(1949)]{Lyapunov49}
  Lyapunov, A. 1949, Ann. Math. Studies, 17

\bibitem[Patsis(2006)]{Patsis06} Patsis, P.A.,
  2006, \mnras, 369, L56

\bibitem[Romero-G\'omez et al.(2006)]{rom06} Romero-G\'omez, M. et al.\ 2006, 
\aap, 453, 39 (Paper I)

\bibitem[Romero-G\'omez et al.(2007)]{rom07} Romero-G\'omez, M. et al.\ 2007,
\aap, 472, 63 (Paper II)

\bibitem[Romero-G\'omez et al.(2008)]{rom08} Romero-G\'omez, M. et al.\ 2008, 
Communications in Nonlinear Science and Numerical Simulations, 
DOI: 10.1016/j.cnsns.2008.07.013 and astro-ph/0807.3832

\bibitem[Rautiainen, Salo \& Laurikainen (2008)]{RautiainenSL08}Rautiainen,
  P., Salo, H., Laurikainen, E. 2008, \mnras, 388, 1803

\bibitem[Schwarz(1979)]{sch79} Schwarz, M.P.\ 1979, Ph.D. Thesis, Australian National University

\bibitem[Schwarz(1981)]{sch81} Schwarz, M.P.\ 1981, \apj, 247, 77

\bibitem[Schwarz(1984)]{sch84} Schwarz, M.P.\ 1984, \mnras, 209, 93

\bibitem[Schwarz(1985)]{sch85a} Schwarz, M.P.\ 1985, \mnras, 212, 677

\bibitem[Toomre(1963)]{too63} Toomre, A.\ 1963, \apjs, 138, 385

\bibitem[Tsoutsis, Efthymiopoulos \& Voglis(2008)]{TsoutsisEV08}
  Tsoutsis, P., Efthymiopoulos, C., Voglis, N. 2008, \mnras, 387, 1264

\bibitem[van den Bergh(1976)]{vdBergh76} van den Bergh, S.\ 1976,
  \apj, 206, 883

\bibitem[Wada \& Koda(2001)]{WadaKoda01} Wada, K., Koda, J. 2001,
  \pasj, 53, 1163 


\end{thebibliography}

\end{document}